\documentclass[letterpaper, 10 pt, conference]{IEEEtran}

\usepackage{textcomp}
\usepackage{graphicx}
\usepackage{caption}
\usepackage{subcaption}
\usepackage{float}

\begin{document}

\pagenumbering{arabic}

\title{SURF-SVM Based Identification and Classification of Gastrointestinal Diseases in Wireless Capsule Endoscopy}
\author{\IEEEauthorblockN{Vanshika Vats\IEEEauthorrefmark{1}, Pooja Goel\IEEEauthorrefmark{2}, Amodini Agarwal\IEEEauthorrefmark{3}, Nidhi Goel\IEEEauthorrefmark{4}\\
\IEEEauthorblockA{Department of Electronics and Communication\\
Indira Gandhi Delhi Technical University for Women\\
Delhi-110006}
Email: \IEEEauthorrefmark{1}vanshika18187@iiitd.ac.in,
\IEEEauthorrefmark{2}goelpooja40@gmail.com,
\IEEEauthorrefmark{3}p18amodinia@iimidr.ac.in,
\IEEEauthorrefmark{4}nidhigoel@igdtuw.ac.in}}
\maketitle
\begin{abstract}
\fontsize{9 pt}{12 pt}
Endoscopy provides a major contribution to the diagnosis of the Gastrointestinal Tract (GIT) diseases. With Colon Endoscopy having its certain limitations, Wireless Capsule Endoscopy is gradually taking over it in the terms of ease and efficiency. WCE is performed with a miniature optical endoscope which is swallowed by the patient and transmits colour images wirelessly during its journey through the GIT, inside the body of the patient. These images are used to implement an effective and computationally efficient approach which aims to detect the abnormal and normal tissues in the GIT automatically, and thus helps in reducing the manual work of the reviewers. The algorithm further aims to classify the diseased tissues into various GIT diseases that are commonly known to be affecting the tract. In this manuscript, the descriptor used for the detection of the interest points is Speeded Up Robust Features (SURF), which uses the colour information contained in the images which is converted to CIELAB space colours for better identification. The features extracted at the interest points are then used to train and test a Support Vector Machine (SVM), so that it automatically classifies the images into normal or abnormal and further detects the specific abnormalities. SVM, along with a few parameters, gives a very high accuracy of 94.58\% while classifying normal and abnormal images and an accuracy of 82.91\% while classifying into multi-class. The present work is an improvement on the previously reported analyses which were only limited to the bi-class classification using this approach.
\end{abstract}


\section{Introduction}
Endoscopy is a way to see inside the human body in a least invasive way. It is a non invasive examination or surgical procedure. This is done using a medical device, endoscope, which is inserted into the interior of a body cavity or a hollow organ . Nowadays, with advancement in technology, we have new technique for endoscopy: Wireless Capsule Endoscopy (WCE). Wireless Capsule Endoscopy is performed with a miniature optical endoscope, which is swallowed by the patient. In the course of an endoscopic procedure, the endoscopic camera produces a video signal which is visualised to a team of doctors which helps them to decide their actions. It transmits colour images wirelessly during its journey inside the body of the patient. 

A study states that it takes a reviewer about 45-90 minutes \cite{time} assessing the video which requires undistracted and intense focus. If this time were to be reduced, more attention would be given to the treatment of the ailment recognised than to the long and tiring manual diagnosis. This makes it essential to develop a method that is primarily made to assist the physician in the screening process. Hence, there was a need to propose some research methods to process and analyse it, either post procedural or in real time. Either way, image processing techniques are often used to pre-process individual video frames and improve their visual quality or performance of the next processing steps.

Most of these problem-tackling solutions were based on textural or colour features. For example, bleeding detection was done based on chromaticity moments \cite{chromaticity}, ulcer detection based on colour rotation in RGB colour space \cite{RGB}, combined ulcer-polyp detection in HSV colour space using Gabor filter \cite{gabor}. But the one thing common in the previously implemented methods is that they were only binary class. The features extracted were only used to classify the video frames into normal or abnormal-whether containing the disease or not. With a diverse disease set available for the gastrointestinal tract, it becomes necessary not only to classify into a normal or abnormal frame, but into further classes of diseases. 

The method proposed in this manuscript aims to analyse the images so as to train an algorithm in a way that it is able to differentiate between the endoscopic images with and without abnormalities. It also aims to classify the images with abnormalities into further classes. This will be useful to the reviewers in increasing the efficiency and reducing the computational time of analysing the images manually. The procedure follows decorrelating the RGB components of the images into a different colour space and applying Speeded Up Robust Features (SURF) algorithm to find out the interest points based on colours. The features are further extracted from those points to be fed to SVM classifier. The process is followed on an entire dataset of images and then classified into normal or abnormal images, and further into the specific diseases. Finally, the accuracy and recall of the classifier trained is found out. 
\vspace{2cm}
\section{Methodology}
Colour is one of the most important aspect in endoscopic image analysis. The colour differences between the abnormality and mucosa attracts the eye of the reviewer of endoscopic images the most, and he/she is then able to make predictions about the diseased abnormality. Texture of the abnormality can then be assessed to differentiate more between the diseases of the similar colour components but this is mostly in the case of High Definition (HD) endoscopy.  
The proposed methodology aims to classify the endoscopic images broadly into two classes, i.e. normal tissue and abnormal tissue. These two classes are further divided into eight subclasses of various diseases. It involves the following steps:
\begin{enumerate}
\item Colour space transformation: The RGB (Red-Green-Blue) colour space, in which the endoscopic images originally are, is converted into a space in which the chromatic components are approximately de-correlated from the luminance components. Such spaces include YCbCr, HSV (Hue-Saturation-Value), CIELab etc \cite{color}.
\item Selecting Region of Interest: The images used are annotated and the region of interest is selected out so that the interest point selection algorithm does not have to search the whole image for the required information. It will look into only that area which the annotation has extracted out. 
\item Interest point selection using SURF: SURF algorithm is applied on the RoI, on the component which is the most suitable for detection. SURF is a local feature detector and descriptor which is an advancement of Scale Invariant Feature Transform (SIFT). It is able to detect the points based on the colour information in the image. 
\item Feature extraction: Features are extracted from each of the interest point selected. The features used to describe a point are taken as the colour component values at that particular point, and the maximum and minimum values of the colour components around the MxN pixel matrix around the interest point. Histograms have high dimensionality and hence are inefficient for a small neighbourhood. The extensive experimentation showed that the max-min features can capture the differences more robustly \cite{dim}. 
\item Classification: The features extracted are classified using a Support Vector Machine (SVM) using supervised learning.
\end{enumerate}
\section{Implementation}
The dataset used for this implementation was obtained from Royal Infirmary of Edinburgh (University Hospital and referral centre for WCE for the southeast of Scotland, UK) which performed a total of 252 WCE procedures with MiroCam (IntroMedic Co., Seoul, South Korea). The rate of image capturing is 3 frames per second (fps) with a resolution of $320*320$ pixels \cite{dim}. The images were classified by experts into i) polypoid lesions,  ii) inflammatory lesions, including mucosal aphthae and ulcers, mucosal erythema, mucosal cobblestone, and luminal stenosis; iii) lymphangiectasias, including chylous cysts, nodular lymphangiectasias, punctuate lymphangiectasias, and iv) vascular lesions, including angiectasias and/or intraluminal bleeding. 

A total of 412 images with different abnormalities consisting of aphthae, intraluminal bleeding, nodular lymphangiectasias, chylous cysts (a subgroup of lymphangiectatic small-bowel lesions), angiectasias with different bleeding potentials, polyps and vascular lesions were used. 452 normal images (without any visible abnormality but including bubbles and opaque luminal fluid) were also used to make the dataset balanced \cite{KID}.

\subsection{Colour Space Transformation}
There is a stark difference between the colour of the abnormality and the rest of the background mucosa. We thus encash this characteristic of the abnormal endoscopic images and use it in the methodology to determine the discrimination between the normal and abnormal endoscopic images. 

The RGB (Red, Green, Blue) endoscopic images are first converted into a colour space in which the components are de-correlated from the luminance component. Such colour spaces may include HSV (Hue, Saturation, Value) and CIELab colour spaces \cite{color}. They are more near to the Human Vision System. The following figures show the colour transformation of an RGB endoscopic image to various colour components of the HSV and CIELab colour models.

\begin{figure}[h]
\begin{subfigure}[b]{\linewidth}
\begin{center}
  \includegraphics[width=0.3\linewidth]{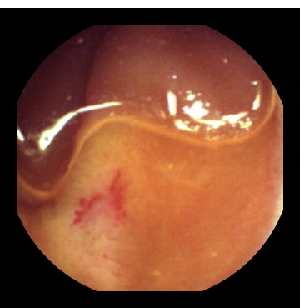}
  \caption{Original Image}
   \end{center}
  \end{subfigure}


\begin{center}
\begin{subfigure}[b]{0.3\linewidth}
  \includegraphics[width=\linewidth]{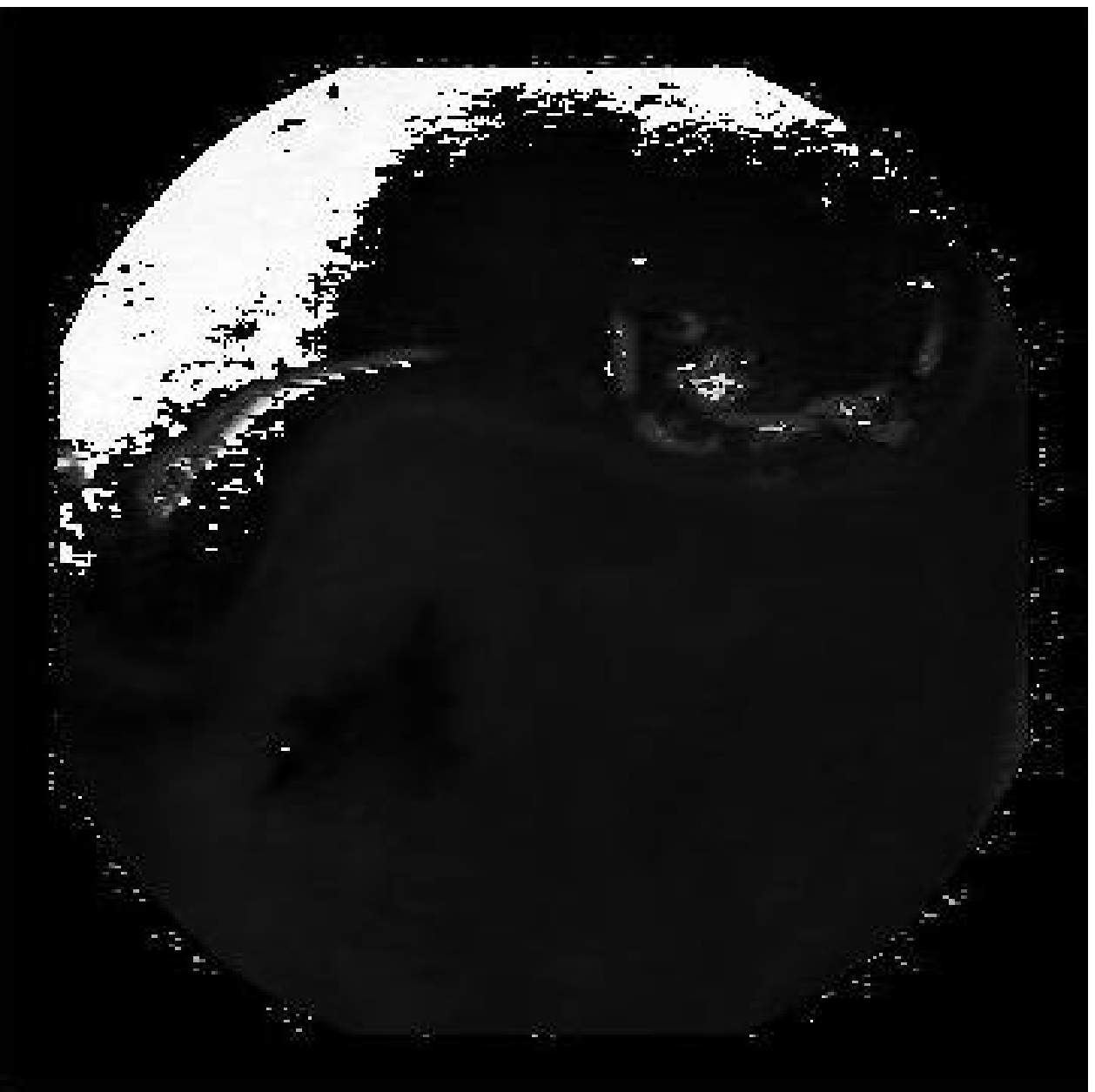}
  \caption{H Component}
  \end{subfigure}
  \hspace{0.1 cm}
  \begin{subfigure}[b]{0.3\linewidth}
  \includegraphics[width=\linewidth]{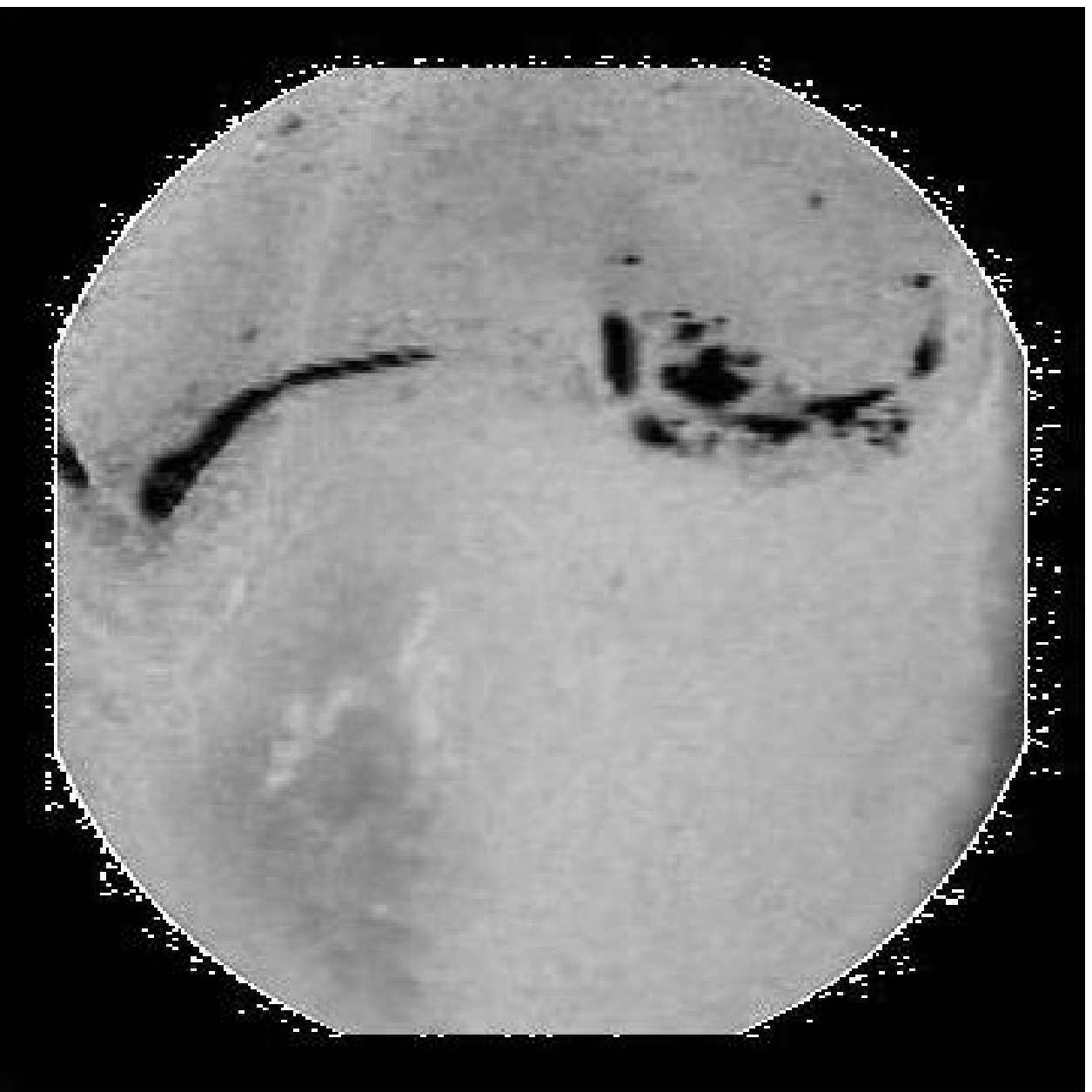}
  \caption{S Component}
   \end{subfigure}
  \hspace{0.1 cm}
  \begin{subfigure}[b]{0.3\linewidth}
 \includegraphics[width=\linewidth]{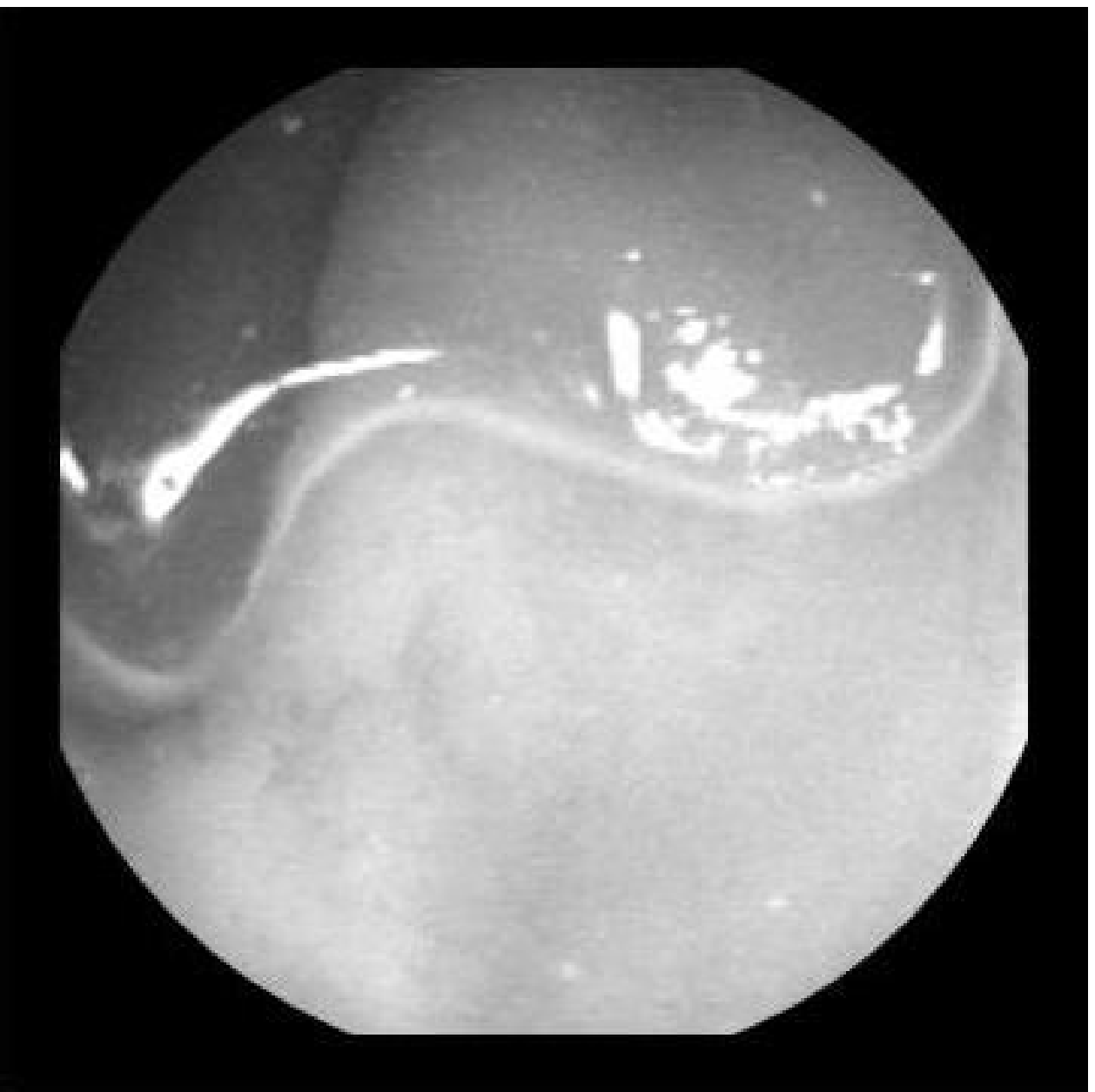}
  \caption{V Component}
   \end{subfigure}
   \end{center}


\begin{center}
\begin{subfigure}[b]{0.3\linewidth}
  \includegraphics[width=\linewidth]{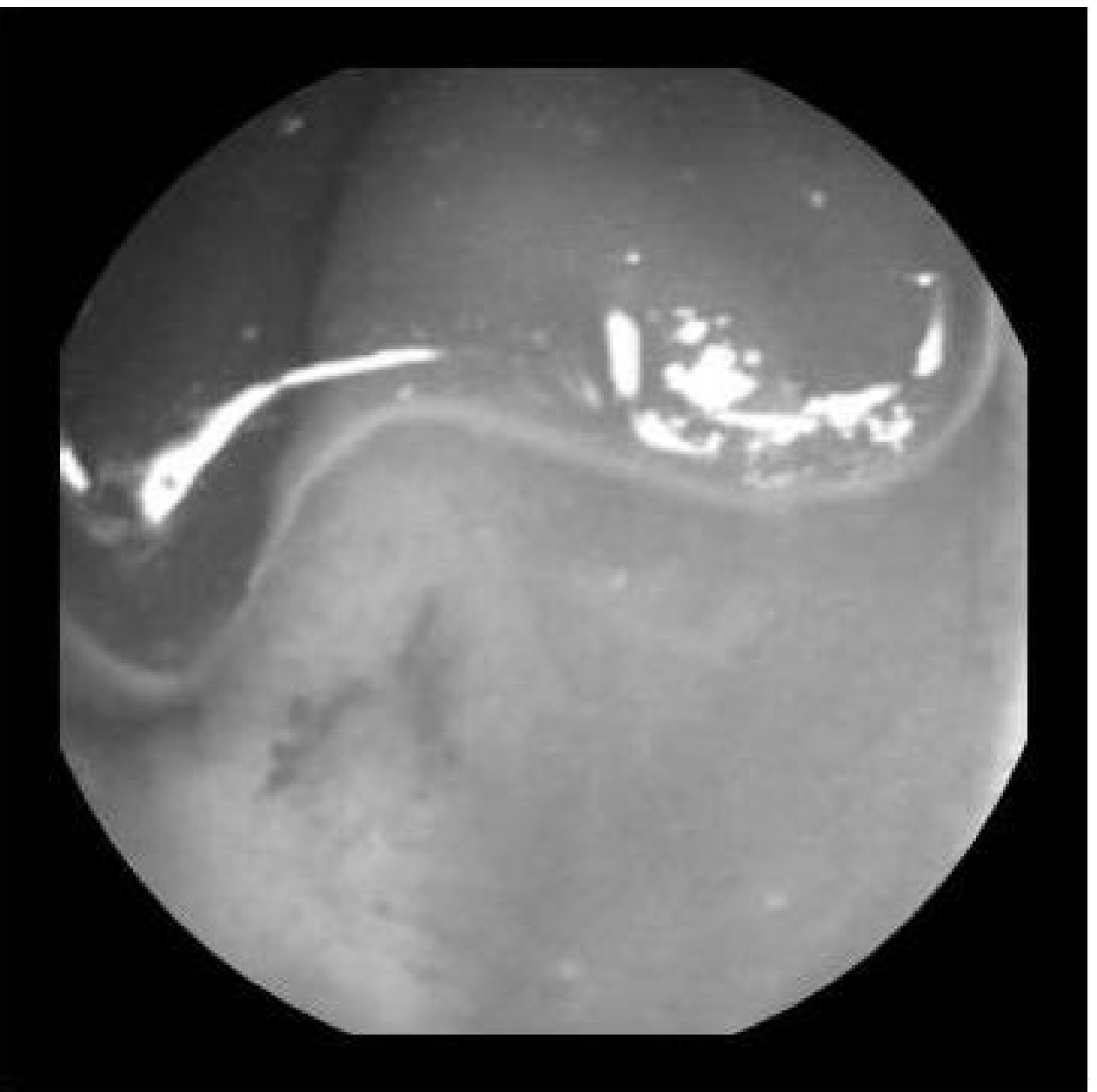}
  \caption{L Component}
  \end{subfigure}
  \hspace{0.1 cm}
  \begin{subfigure}[b]{0.3\linewidth}
  \includegraphics[width=\linewidth]{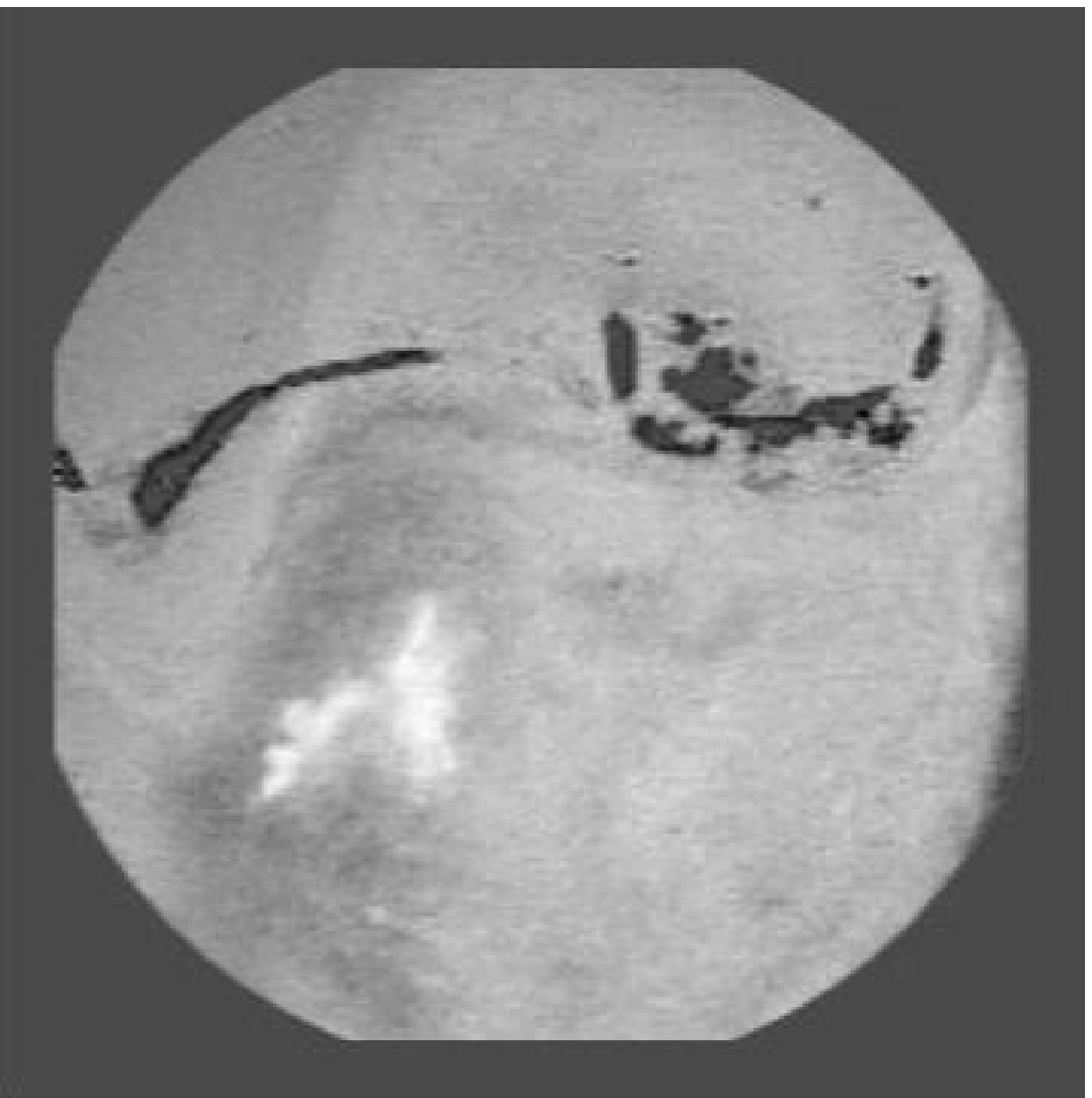}
  \caption{a Component}
  \end{subfigure}
  \hspace{0.1 cm}
  \begin{subfigure}[b]{0.3\linewidth}
  \includegraphics[width=\linewidth]{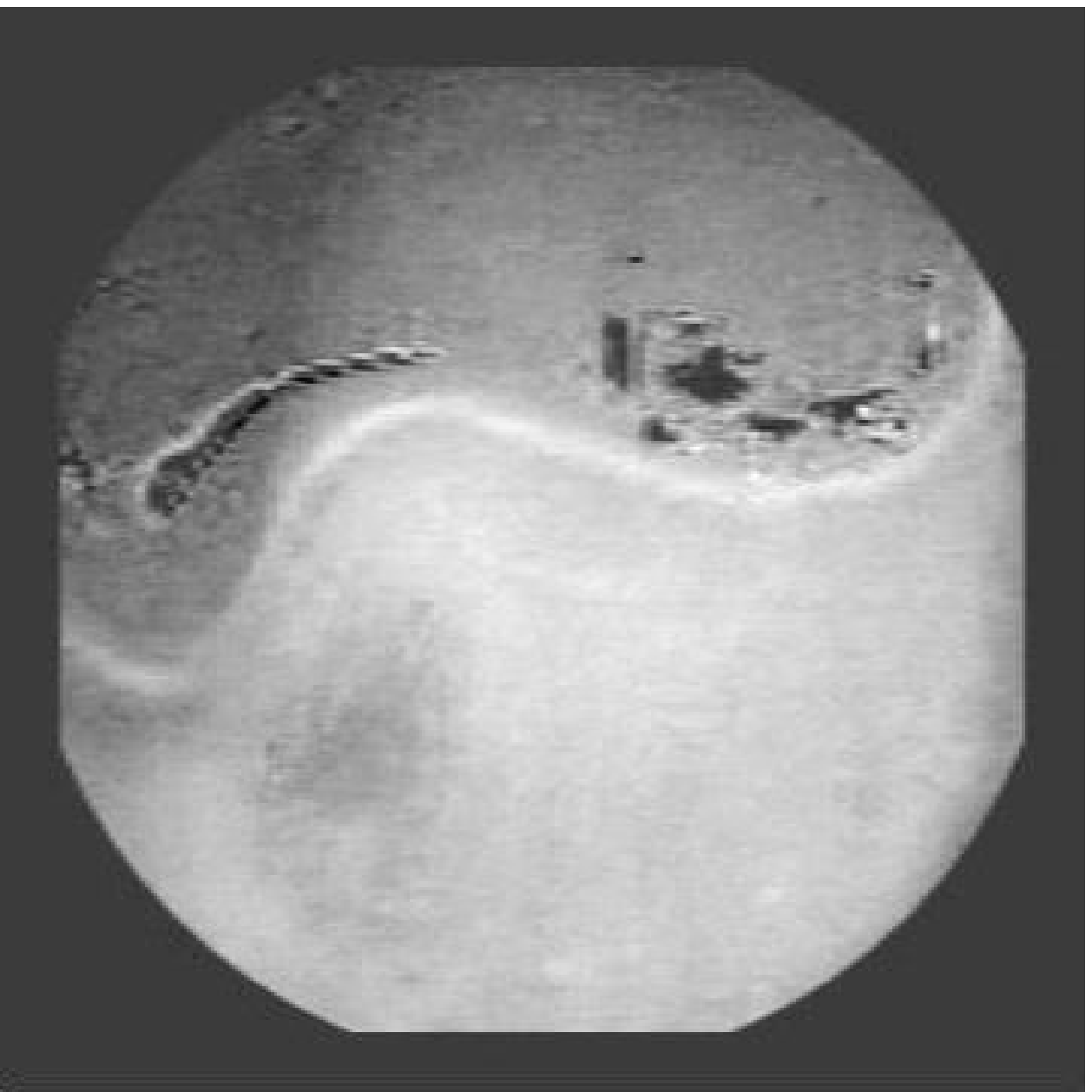}
  \caption{b Component}
  \end{subfigure}
  \end{center}
  \caption{Colour space transformation into HSV and CIELab}
\end{figure}

\subsection{Selection of Region of Interest}
Image annotation refers to marking of an image in various shapes according to our needs by an algorithm so as to carry out a particular operation/operations in the specified region. 

The images of the dataset were graphically annotated to pixel level details by the experts using the Ratsnake Annotation Tool \cite{rat}. These annotations were used to find out the Region of Interest, such that the following algorithm search for the points in the specified areas, and not the whole image, thus effectively reducing the computational time.

\begin{figure}[h]
\begin{center}
\begin{subfigure}[b]{0.35\linewidth}
  \includegraphics[width=\linewidth]{angioectasia-P1-4}
  \caption{Original Image}
  \end{subfigure}
  \hspace{0.6 cm}
  \begin{subfigure}[b]{0.35\linewidth}
  \includegraphics[width=\linewidth]{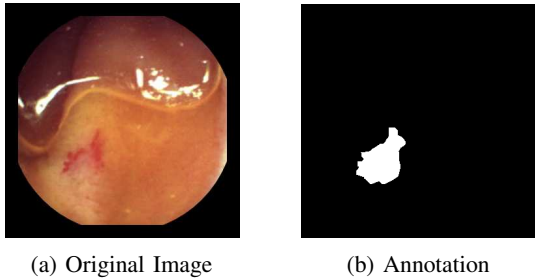}
  \caption{Annotation}
  \end{subfigure}
  
 \vspace{0.5 cm}
 
  \begin{subfigure}[b]{0.35\linewidth}
  \includegraphics[width=\linewidth]{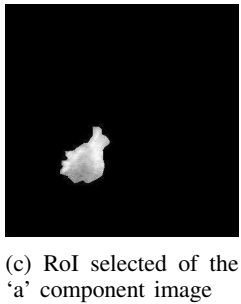}
  \caption{RoI selected of the `a' component image}
  \end{subfigure}
\end{center}
  \caption{Selection of Region of Interest}
\end{figure}

\subsection{Selection of interest points using SURF}
To classify $(412+452)*320*320 \approx 8.8*10^{7}$ pixels is a challenging task. To make this process computationally efficient, the number of pixels of interest are significantly reduced by applying the SURF algorithm on the image dataset. 

The Speeded Up Robust Features (SURF) algorithm determines the points where the direction of the boundary changes abruptly \cite{SURF}. It is scale invariant and computes the interest points by the convolutions of input image with filters at multiple scales and fast approximation of Hessian Matrix. This finding of specific interest points reduces the pixels to be classified significantly. But a minimum number of points should still be specified so as not to miss even a small abnormality. 

The interest points are selected automatically based solely on the colour information. The algorithm was applied on each of the components specified above, i.e. Hue, Saturation and Value in HSV and L, a and b component in CIELab colour space. It was found out that the component `a' gave the best results among all \cite{dim}. It gave the maximum number of correct interest points at most of the SURF threshold values defined. Hence, the further work was carried out only in the `a' component of the images.

\begin{figure}[h]
\begin{center}
  \includegraphics[width=\linewidth]{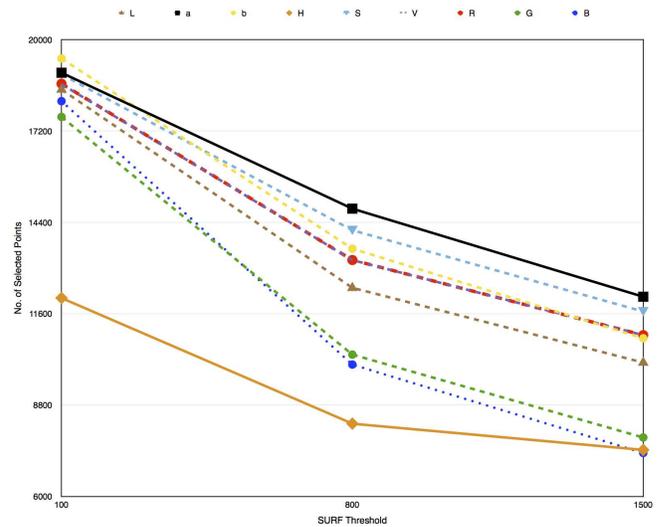}
  \end{center}
 \caption{Performance Comparison}
\end{figure}

Thus, the interest points detected using SURF algorithm on the `a' component of the CIELab colour space can be represented as the centre of the circular areas as shown in Fig (4).

\begin{figure}[h]
\begin{center}
  \includegraphics[width=\linewidth]{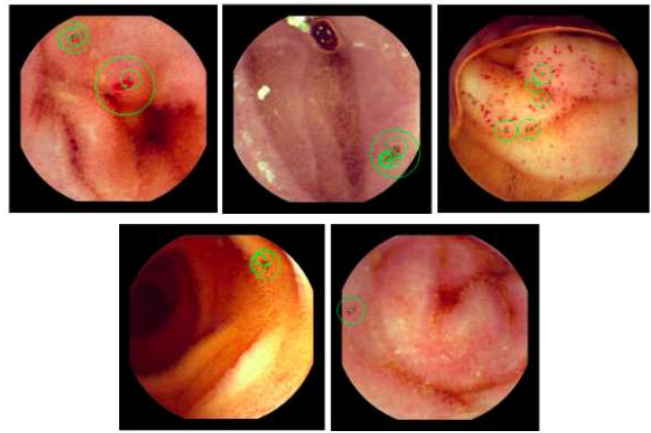}
  \end{center}
 \caption{Interest points selected using SURF on `a' component}
\end{figure}

\subsection{Feature Extraction}
The next and the most important task is to extract some features from the image that can be used to find out a pattern and thus train the algorithm. These features define the characteristics of the interest points selected by the SURF algorithm. 

The colour information was once again used to find out the features. The features were either selected by the pixel assessment or the assessment of the maximum-minimum values using statistical features. The feature set was formed using the `L' (illuminance), `a' (chromatic Green-Red), `b' (chromatic Blue-Yellow) values on the original image at the points detected using the SURF algorithm. The size of this neighbourhood was decided to be $36*36$ pixels based on the preliminary classification. 
The final feature set includes a total of 9 features:
\(S=\{L, a, b, Max_L, Min_L, Max_a, Min_a, Max_b, Min_b\}\), where $Max_x$ and $Min_x$ represent the maximum and minimum values, respectively, of a colour component `x' in the $36*36$ neighbourhood of each interest point. 

\begin{figure}[h]
\begin{center}
  \includegraphics[width=\linewidth]{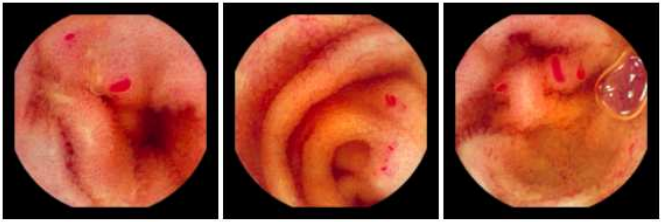}
  \end{center}
 \caption{Original Images}
\end{figure}
\begin{figure}[h]
\begin{center}
  \includegraphics[width=\linewidth]{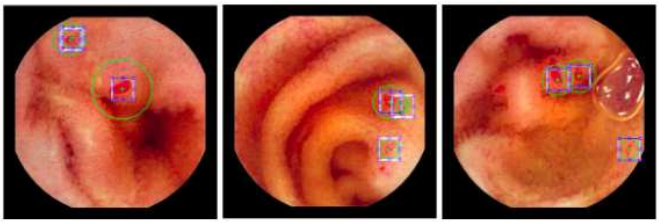}
  \end{center}
 \caption{Regions with interest points and selected neighbourhood}
\end{figure}

The feature extraction algorithm is run over the entire dataset and a table is made containing the values of set S, for the SURF detected interest points. These features are used to feed the classifier.

\subsection{Classification}
The data set formed by the extraction of feature values is fed to the SVM classifier. The training and the test data is divided into the ratio of 0.75:0.25. Feature scaling is done by the SVM Classifier followed by training the classifier to generate the predicting variable. A Confusion Matrix is generated to describe the performance of the prediction model. For multidimensionality reduction, feature extraction is done using Kernel PCA, which allows this multidimensional data (here, features extracted) to get reduced to bi-dimensional data for better representational purposes. The principal components which are related non-linearly to the input space are found by the Kernal PCA. This is done by applying PCA in the space produced by the nonlinear mapping, where the low-dimensional latent structure is, hopefully, easier to discover. It therefore helps in data compression and thus reduced storage space and computation time. 

The data is classified using both linear kernel and Radial Basis Function (RBF) kernel. RBF is helpful where it is difficult to decide a linear decision boundary for a data set. It can automatically decide a non-linear one.  

\subsection{Structure}

The flowchart demonstrating the structure of the methodology is shown below:

\vspace{1 cm}
\begin{figure}[h]
\begin{center}
  \includegraphics[width=\linewidth]{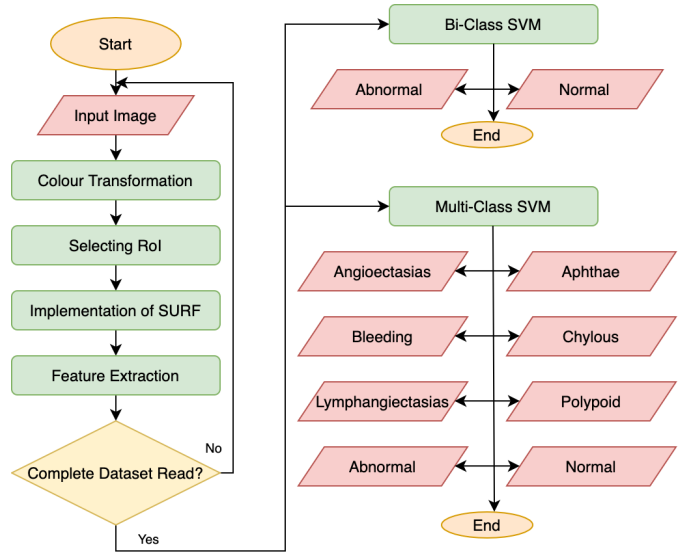}
  \end{center}
 \caption{Flowchart explaining the structure of the methodology}
\end{figure}

\section{Results}
The results of bi-class and multi-class classification are tabulated in Table 1. It can be seen that using a RBF kernel gives a significantly higher accuracy than linear kernel for both bi-class (94.58\%) and multi-class (82.91\%). The F1 score is also considered which gives a balance between precision and recall as it is important in medical testing to know and reduce the percentage of false negatives, i.e. a disease actually present should not be diagnosed wrongly as not present. 

\vspace{0.2 cm}
\begin{table}[H]
\caption{Classification Results}
 \begin{tabular}{| c | c | c | c | c |}
 \hline
 &\multicolumn{2}{|c|}{Without Kernel PCA}&\multicolumn{2}{c|}{With Kernel PCA}\\
 \hline
 & F1 Score & Accuracy(\%) & F1 score & Accuracy(\%)\\
 \hline
 \textbf{Bi-class} & \multicolumn{4}{c|}{}\\
 \hline
 Linear & 0.88 & 87.75 & 0.85 & 84.91\\
 \hline
 RBF & \textbf{0.95} & \textbf{94.58} & 0.85 & 85.33\\
 \hline
 \textbf{Multi-class} & \multicolumn{4}{c|}{}\\
 \hline
 Linear & 0.71 & 75.08 & 0.53 & 62.25\\
 \hline
 RBF & \textbf{0.82} & \textbf{82.91} & 0.53 & 62.41\\
 \hline
 \end{tabular}
 \end{table}

\vspace{1 cm}

 \section{Conclusion and Discussion}
 The proposed approach converts the images into numerical data and then processes it which makes the processing computationally efficient, without compromising with the accuracy. The manuscript displays a state of the art approach which works well in the scenarios which cannot afford high computational power and time. SURF thus helps to extract out features and process the data in the numerical form in a relatively lower time and without needing high computational cost and time. SVM works as an excellent classifier to segregate the images into two broad classes, i.e. normal and abnormal and further classifies it into 7 classes, namely, Angioectasias, Apthae, Chylous Cysts, Nodular Lymphangiectasias, Polypoids, Vascular lesions, and Bleeding in the tissue. This is an improvement and extension on the previous works with SURF and SVM as the literature in this field suggests only bi-class classification in normal and abnormal tissues. Thus, the manuscript largely focusses on a simple, efficient and yet a highly accurate approach which classifies a few of the most common GIT diseases into their respective classes.\\
Our proposed approach delves further into the classification and performs acceptably well on the 7 class classification. The analysis has been done on the available dataset and the accuracy using SURF and SVM is found out to be of 94.58\% for bi-class classification and 82.91\% for multi-class classification. Thus, the algorithm carries out a rather good trade-off between the cost of processing and an extremely high accuracy. The drop in the accuracies between the bi-class and multi-class classification owes to the fact that the available data was somewhat correlated even after processing it to get the maximum information out of it. This made the classifier to distinguish the data a little inaccurately which explains the drop. Also, using Kernel PCA for dimensionality reduction along with SVM classifier reduces the accuracy too. This happens because Kernel PCA uses non-linear multidimensionality reduction which analyses the pre-fed features and combines them to make new ones. Since images and their features extracted from the dataset are somewhat correlated and show low variance. Kernel PCA makes the data lose some spatial information which might have been important for classification, which reduces the accuracy. Thus, SVM can be used standalone, without Kernel PCA, for better results. Further works can be done on procuring a more accurate and uncorrelated dataset and processing it in a more efficient manner.

\bibliographystyle{ieeetr}
\bibliography{endoscopy_new}

\begin{thebibliography}{10}

\bibitem{time}
S.~K. Lo, ``How should we do capsule reading?,'' {\em Techniques in
  Gastrointestinal Endoscopy}, vol.~8, no.~4, pp.~146--148, 2006.

\bibitem{chromaticity}
B.Li and M.-H. Meng, ``Computer-aided detection of bleeding regions for capsule
  endoscopy images,'' {\em IEEE Transactions on Biomedical Engineering},
  vol.~56, no.~4, pp.~1032--1039, 2009.

\bibitem{RGB}
V.~S. Charisis, C.~Katsimerou, L.~J. Hadjileontiadis, C.~N. Liatsos, and G.~D.
  Sergiadis, ``Computer-aided capsule endoscopy images evaluation based on
  color rotation and texture features: An educational tool to physicians,''
  {\em 2013 IEEE 26th International Symposium on Computer-Based Medical Systems
  (CBMS)}, pp.~203--208, 2013.

\bibitem{gabor}
P.~Szczypinski, A.~Klepaczko, M.~Pazurek, , and P.~Daniel, ``Texture and color
  based image segmentation and pathology detection in capsule endoscopy
  videos,'' {\em Computer Methods and Programs in Biomedicine}, vol.~113,
  no.~1, pp.~396--411, 2014.

\bibitem{color}
G.~Wyszecki and W.~S. Stiles, {\em Color science}, vol.~8.
\newblock Wiley New York, 1982.

\bibitem{dim}
D.~K. Iakovidis and A.~Koulaouzidis, ``Automatic lesion detection in wireless
  capsule endoscopy - a simple solution for a complex problem,'' in {\em
  International Conference on Image Processing}, 2014.

\bibitem{KID}
Koulaouzidis and D.~Iakovidis, ``Kid: Koulaouzidis-iakovidis database for
  capsule endoscopy,'' 2015.

\bibitem{rat}
D.~Iakovidis, T.~Goudas, C.~Smailis, and I.~Maglogiannis, ``Ratsnake: A
  versatile image annotation tool with application to computer-aided
  diagnosis,'' {\em The Scientific World Journal}, vol.~2014, 2014.

\bibitem{SURF}
H.~Bay, T.~Tuytelaars, , and L.~V. Gool, ``Surf: Speeded up robust features,''
  {\em European Conference on Computer Vision}, pp.~404--417, 2006.

\bibitem{new3}
M.~Souaidi, S.~Charfi, A.~A. Abdelouahad, and M.~E. Ansari, ``New features for
  wireless capsule endoscopy polyp detection,'' in {\em 2018 International
  Conference on Intelligent Systems and Computer Vision (ISCV)}, pp.~1--6,
  2018.

\bibitem{new2}
C.~P. Sindhu and V.~Valsan, ``Automatic detection of colonic polyps and tumor
  in wireless capsule endoscopy images using hybrid patch extraction and
  supervised classification,'' in {\em 2017 International Conference on
  Innovations in Information, Embedded and Communication Systems (ICIIECS)},
  pp.~1--5, 2017.

\bibitem{new}
Y.~Yuan, B.~Li, and M.~Q. Meng, ``Wce abnormality detection based on saliency
  and adaptive locality-constrained linear coding,'' {\em IEEE Transactions on
  Automation Science and Engineering}, vol.~14, no.~1, pp.~149--159, 2017.

\bibitem{26}
C.~J. Burges, {\em A tutorial on support vector machines for pattern
  recognition}, vol.~2.
\newblock 1998.

\bibitem{dimension}
A.~Ghodsi, {\em Dimensionality Reduction A Short Tutorial}.
\newblock Department of Statistics and Actuarial Science, University of
  Waterloo, 2006.

\bibitem{kernel}
Q.~Wang, ``Kernel principal component analysis and its applications in face
  recognition and active shape models,'' {\em CoRR}, vol.~abs/1207.3538, 2012.

\bibitem{bright}
D.~R. Martin, C.~C. Fowlkes, and J.~Malik, ``Learning to detect natural image
  boundaries using local brightness, color, and texture cues,'' {\em IEEE
  Transactions}, vol.~26, no.~530-549, 2014.

\bibitem{SIFT}
P.~M. Panchal, S.~R. Panchal, and S.~K. Shah, ``A comparison of sift and
  surf,'' {\em International Journal of Innovative Research in Computer and
  Communication Engineering}, vol.~1, April 2013.

\bibitem{extra2}
S.Georgakopoulos, D.K.Iakovidis, M.Vasilakakis, and V.P.Plagianakos, ``Weakly-
  supervised convolutional learning for detection of inflammatory
  gastrointesti- nal lesions,'' in {\em IEEE International Conference on
  Imaging Systems and Techniques (IST)}, (Chania, Greece), pp.~510--514, 2016.

\bibitem{extra}
M.Vasilakakis, D.K.Iakovids, E.Spyrou, and A.Koulaouzidis, ``Weakly-supervised
  lesion detection in video capsule endoscopy based on a bag-of-colour features
  model,'' in {\em Lecture Notes in Computer Science}, pp.~97--104, MICCAI
  Workshop on Computer Assisted and Robotic Endoscopy (CARE), 2016.

\bibitem{Iakovidis:2015aa}
D.~K. Iakovidis, Koulaouzidis, and Anastasios, ``Software for enhanced video
  capsule endoscopy: challenges for essential progress,'' {\em Nat Rev
  Gastroenterol Hepatol}, vol.~12, no.~3, pp.~172--86, 2015.

\bibitem{new2020}
S.~Ali, F.~Zhou, B.~Braden, A.~Bailey, S.~Yang, G.~Cheng, P.~Zhang, X.~Li,
  M.~Kayser, R.~D. Soberanis-Mukul, S.~Albarqouni, X.~Wang, C.~Wang,
  S.~Watanabe, I.~Oksuz, Q.~Ning, S.~Yang, M.~A. Khan, X.~W. Gao, S.~Realdon,
  M.~Loshchenov, J.~A. Schnabel, J.~E. East, G.~Wagnieres, V.~B. Loschenov,
  E.~Grisan, C.~Daul, W.~Blondel, and J.~Rittscher, ``An objective comparison
  of detection and segmentation algorithms for artefacts in clinical
  endoscopy,'' {\em Scientific Reports}, vol.~10, no.~1, p.~2748, 2020.

\end{thebibliography}
\nocite{*}

\end{document}